\documentclass[nobibnotes,preprintnumbers,aps,prl,superscriptaddress,showpacs,twocolumn,twoside,sort&compress]{revtex4}				 %
\usepackage{graphicx}		%
\usepackage{epstopdf}		% added by user										 
\usepackage{titlesec}																											     %
\usepackage{fancyheadings}
\usepackage{color}																							 				 %
\begin{document}																													 %
%%%%%%%%%%%%%%%%%%%%%%%%%%%%																										 %
%																																	 %	
\preprint{{Vol.XXX (201X) ~~~~~~~~~~~~~~~~~~~~~~~~~~~~~~~~~~~~~~~~~~~~~~~~~~~~ {\it CSMAG`16}										  ~~~~~~~~~~~~~~~~~~~~~~~~~~~~~~~~~~~~~~~~~~~~~~~~~~~~~~~~~~~~ No.X~~~~}}																 %			
\vspace*{-0.3cm}																													 %
\preprint{\rule{\textwidth}{0.5pt}}																											 \vspace*{0.3cm}																														 %
%%%%%%%%%%%%%%%%%%%%%%%%%%%%%%%%%%%%%%%%%%%%%%%%%%%%%%%%%%%%%%%%%%%%%%%%%%%%%%%%%%%%%%%%%%%%%%%%%%%%%%%%%%%%%%%%%%%%%%%%%%%%%%%%%%%%%%

\title{Enhanced magnetocaloric effect due to selective dilution in a triangular Ising antiferromagnet}

% author.  The \affiliation command should follow the
\author{M. Borovsk\'{y}}
\thanks{corresponding author; e-mail: borovsky.michal@gmail.com}
\author{M. \v{Z}ukovi\v{c}}
\affiliation{Institute of Physics, Faculty of Sciences, P. J. \v{S}af\'{a}rik University, Park Angelinum 9, 040 01 Ko\v{s}ice, Slovakia}

% abstract
\begin{abstract}
We employ an effective-field theory with correlations in order to study a magnetocaloric effect on a triangular Ising antiferromagnet, which is selectively diluted by non-magnetic impurities on one of the three sublattices. Such a dilution generally relieves massive degeneracy in our system and therefore the ground-state entropy diminishes and the magnetocaloric effect weakens at low temperatures. However, at relatively higher temperatures we can observe significantly enhanced negative isothermal entropy changes for the sublattice concentration $p_{\mathrm{A}} = 0.8$.
%Additional Monte Carlo simulations confirmed that a selective dilution can indeed decrease the residual entropy, which is a consequence of large degeneracy in the system. But unfortunately, these simulations also disprove the existence of enhanced magnetocaloric behavior and thus we consider this result to be an artifact of the effective-field theory.
\end{abstract}


\maketitle

\section{Introduction}
An Ising antiferromagnet on a triangular lattice is the most rudimentary example of a geometrically frustrated system. With such a topology of the lattice it is impossible to create a microstate where all couplings have simultaneously the same minimal contribution to the internal energy and this results in an infinite-fold degeneracy at zero temperature point and no long-range ordering as has been shown in Wannier's work \cite{ref1}. This translates to a large residual entropy and for this system it was calculated to be $S_0 = 0.323066 Nk_B$.

In our previous work \cite{ref2}, we examined a joint effect of an external magnetic field and a selective dilution with non-magnetic impurities or vacancies, which were infused to only one of the three sublattices. Such perturbations help to relieve degeneracy and even induce long-range ordering in the system for a small degree of the dilution. Moreover, relieving the massive degeneracy is also related to considerable negative isothermal entropy changes \cite{ref3}, which are relevant in a magnetocaloric effect.

In this study, we focus on the effect of a selective dilution on magnetocaloric properties of the system at moderate external magnetic fields within the framework of an effective-field theory with correlations \cite{ref4}.
\section{Effective-field theory}
The diluted Ising model in an external magnetic field can be described by the Hamiltonian
\begin{equation}
H = -J \sum\limits_{\left\langle i,j\right\rangle} \xi_i \xi_j S_i S_j - h \sum\limits_{i} \xi_i S_i ,
\label{eq:1}
\end{equation}
where $S_i = \pm 1$ are Ising spin variables, $J<0$ is an antiferromagnetic exchange interaction, $h$ is the external magnetic field, $\left\langle i,j\right\rangle$ denotes the summation over all nearest neighbor couplings in the system and $\xi_i = 1$ or $0$ are quenched, uncorrelated random variables, where $\xi_i = 1$ represents occupancy of the site $i$ with a magnetic atom with the probability $p$. The quantity $p$ also represents the mean concentration of magnetic atoms. In our selectively diluted system this concentration is sublattice-dependent and we are considering only one sublattice (lets say A) out of three to be diluted, i.e. $0 \leq p_{\mathrm{A}} \leq 1, p_{\mathrm{B}} = p_{\mathrm{C}} = 1$.

Following the same procedure for the effective-field theory (EFT) with correlations as in our previous work \cite{ref2}, we can derive a system of coupled nonlinear equations for the sublattice magnetizations per spin $m_{\mathrm{A}}, m_{\mathrm{B}}$ and $m_{\mathrm{C}}$ in the form
\begin{equation}
\begin{array}{l}
		\label{eq:2}
				m_{\mathrm{A}} = p_{\mathrm{A}} \left( a + b m_{\mathrm{B}} \right)^3 \left( a + b m_{\mathrm{C}} \right)^3 \tanh \left[ \beta\left( x + h \right) \right] |_{x=0}, \\
				m_{\mathrm{B}} = \left( a_{\mathrm{A}} + b m_{\mathrm{A}} \right)^3 \left( a+ b m_{\mathrm{C}} \right)^3 \tanh \left[ \beta\left( x + h \right) \right] |_{x=0}, \\
				m_{\mathrm{C}} = \left( a_{\mathrm{A}} + b m_{\mathrm{A}} \right)^3 \left( a + b m_{\mathrm{B}} \right)^3 \tanh \left[ \beta\left( x + h \right) \right] |_{x=0},
\end{array}
\end{equation}
where $a_{\mathrm{A}} = 1 - p_{\mathrm{A}} + p_{\mathrm{A}}\cosh(JD)$, $a = \cosh(JD)$, $b = \sinh(JD)$, $\beta = (k_BT)^{-1}$ is the inverse temperature, $k_B$ is the Boltzmann constant and $D = \partial / \partial x$ is the differential operator. The total magnetization per site $m$ is defined as the mean value of sublattice magnetizations $m = (m_{\mathrm{A}} + m_{\mathrm{B}} + m_{\mathrm{C}})/3$. The main observable quantity that determines the degree of the magnetocaloric effect is an isothermal entropy change from a zero-field point, which can be calculated as follows
\begin{equation}
\frac{\Delta S}{Nk_B} = \int\limits^{h}_{0} \frac{\partial m}{\partial T} dh ,
\label{eq:3}
\end{equation}
where $N$ is the number of spin sites in the lattice. The temperature derivative of the total magnetization per site $\partial m / \partial T$ represents the mean value of the temperature derivatives of the sublattice magnetizations $\partial m_{\mathrm{X}} / \partial T$ ($\mathrm{X} = \mathrm{A}, \mathrm{B}$ or $\mathrm{C}$). The latter can be evaluated analytically by differentiating the system of equations (\ref{eq:2}) with respect to the temperature $T$. This leads to the system of linear equations for $\partial m_{\mathrm{X}} / \partial T$, which is easy to solve numerically.

\section{Results and Discussion}

\begin{figure}[h!]
\includegraphics[width=0.76\columnwidth]{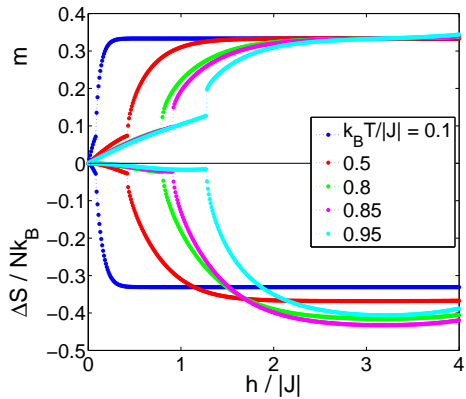}
\caption{Field dependencies of the total magnetization per site $m$ and isothermal entropy changes $\Delta S / Nk_B$ for a non-diluted case ($p_{\mathrm{A}} = 1$) for different temperature values.}
\label{fig1}
\end{figure}

Let us first examine the non-diluted system ($p_\mathrm{A} = p_\mathrm{B} = p_\mathrm{C} = 1$). In Fig. \ref{fig1} we plotted the total magnetization and entropy changes as functions of the external magnetic field for several selected temperature values. As we can see for the lowest displayed temperature of $k_BT/|J| = 0.1$ the entropy change drops drastically fast with a relatively small change of the field at the onset of the $1/3$-magnetization plateau, which is accompanied with the first-order phase transition to the ferrimagnetic phase. Then the entropy change reaches the minimal value of $\Delta S = -0.3309 Nk_B$, which, as expected, is fairly close to the negative value of the residual ($T = 0$) entropy $S_0$ calculated by Wannier \cite{ref1}. When we moderately heat up the system, the decrease of $\Delta S$ is less steep but deeper, which can strengthen the magnetocaloric effect, but at a cost of shifting the minimum to higher fields. This trend continues up to $k_BT / |J| \approx 0.85$, where we observed the minimum value of $-0.4325 Nk_B$ at the field $h/|J| = 3.17$. Further increase of the temperature results in smaller absolute value of $\Delta S$.

\begin{figure}[h!]
\includegraphics[width=0.76\columnwidth]{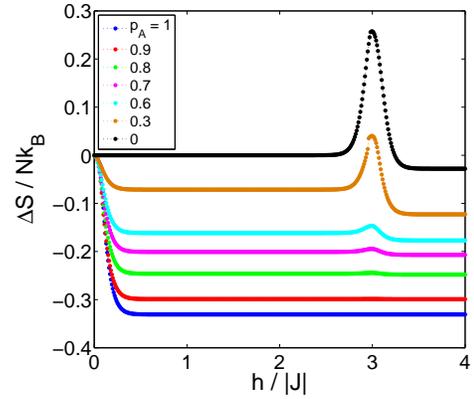}
\caption{Field dependencies of and isothermal entropy changes $\Delta S / Nk_B$ at $k_BT / |J| = 0.1$ for different degrees of the concentration $p_{\mathrm{A}}$ of magnetic atoms on the sublattice A.}
\label{fig2}
\end{figure}

When we introduce the selective dilution to our system, the influence of geometrical frustration is partially diminished. It means that this dilution reduces a degeneracy and consequently the residual entropy of the ground state. We can observe this behavior in Fig. \ref{fig2}, where we plotted the entropy changes for several values of the concentration $p_{\mathrm{A}}$ at $k_BT / |J| = 0.1$, which is near the ground state. We can also see that the case with $p_{\mathrm{A}} = 0$, which corresponds to the non-frustrated Ising antiferromagnet on a honeycomb lattice (see Fig. 1 in Ref. \cite{ref2} for clarity), has a long regime of $\Delta S = 0$. This happens due to the existence two-fold degenerate antiferromagnetic phase, which is also present at $h/|J| = 0$, and thus the entropy has a same value at both zero and non-zero fields.

\begin{figure}[h!]
\includegraphics[width=0.76\columnwidth]{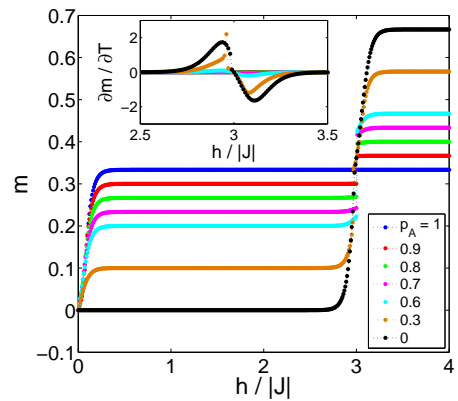}
\caption{Field dependencies of the total magnetization per site $m$ at $k_BT / |J| = 0.1$ for different degrees of the concentration $p_{\mathrm{A}}$. The inset contains dependencies of temperature derivatives of total magnetization $\partial m / \partial T$ near the $h/|J| = 3$ for the corresponding values of $p_{\mathrm{A}}$.}
\label{fig3}
\end{figure}

Another interesting feature, that is visible in Fig. \ref{fig2}, is the existence of local maxima of $\Delta S$ at the field $h/|J| = 3$, which are more pronounced with the decreasing value of concentration $p_A$. The origin of these maxima stems from a fact, that a phase transition occurs in the selectively diluted system at the zero temperature (see Ref. \cite{ref2}). For field values $h/|J| \in (0,3)$ the system displays the ferrimagnetic phase Fi${}_1$ with only one non-diluted sublattice aligned antiparallel to the field, i.e. $(m_{\mathrm{A}}, m_{\mathrm{B}}, m_{\mathrm{C}}) = (p_{\mathrm{A}}, 1, -1)$. However, for $h/|J| \in (3,6)$ a different ferrimagnetic ordering Fi${}_2$ is energetically favorable with the diluted sublattice opposing the field direction, i.e. $(m_{\mathrm{A}}, m_{\mathrm{B}}, m_{\mathrm{C}}) = (-p_{\mathrm{A}}, 1, 1)$. Consequently, these two phases create two plateaus in the magnetization process $m(h)$ with values $p_{\mathrm{A}}/3$ and $(2-p_{\mathrm{A}})/3$ as we can see in Fig. \ref{fig3}. 
The thermal fluctuations at finite temperatures tend to smooth the magnetization jump at $h/|J| = 3$ and thus large changes in the magnetization curves $\partial m / \partial T$ are observed at low temperatures, which are positive for $h/|J| < 3$ and negative for $h/|J| > 3$ (see the inset in Fig. \ref{fig3}). Such a oscillation-like behaviour of $\partial m / \partial T$ yields the existence of $\Delta S$ maxima at $h/|J| = 3$ at low temperatures. Moreover, the distance between two plateaus $\Delta m = 2(1-p_{\mathrm{A}})/3$ at $T=0$ is increasing with the dilution and thus larger magnetization changes and higher maxima of $\Delta S$ are present for smaller $p_{\mathrm{A}}$ values.
\begin{figure}[h!]
\includegraphics[width=0.76\columnwidth]{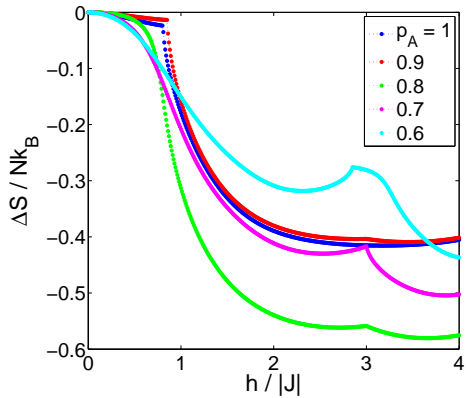}
\caption{Field dependencies of isothermal entropy changes $\Delta S / Nk_B$ at $k_BT / |J| = 0.8$ for different degrees of the concentration $p_{\mathrm{A}}$ of magnetic atoms on the sublattice A.}
\label{fig4}
\end{figure}

When we moderately increase the temperature, the entropy changes do not necessarily diminish with the dilution, but for several values of the concentration $p_{\mathrm{A}}$ we can see the opposite effect with enhanced magnetocaloric properties (see Fig. \ref{fig4}). Specifically, for the case with $p_{\mathrm{A}} = 0.8$ the negative entropy changes extend to $\Delta S = -0.5617 Nk_B$ at $h/|J| = 2.71$ and $k_BT / |J| = 0.8$. 
\begin{figure}[h!]
\includegraphics[width=0.76\columnwidth]{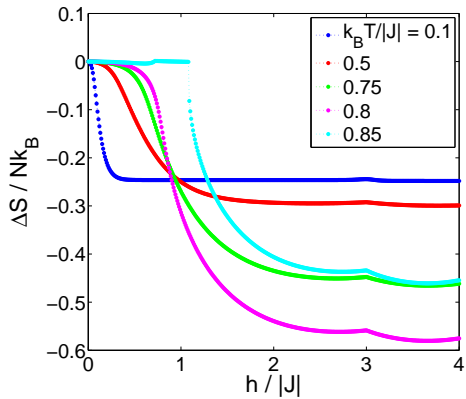}
\caption{Field dependencies of isothermal entropy changes $\Delta S / Nk_B$ for a selectively diluted case with the concentration $p_{\mathrm{A}} = 0.8$ for different temperature values.} 
\label{fig5}
\end{figure}
We can also observe for diluted cases, that the local maximum arises in $\Delta S(h)$ at moderate fields, which has the same nature as in Fig. 2. Moreover, for $h/|J| > 3$ the another minima occurs in $\Delta S(h)$, which shifts even further to negative values, e.g. for $p_{\mathrm{A}} = 0.8$ the minimum of $\Delta S = -0.5804 Nk_B$ is present at $h/|J| = 3.66$. This behavior is also demonstrated in Fig. \ref{fig5}, where we displayed the isothermal entropy changes for a case with fixed $p_{\mathrm{A}} = 0.8$ and several chosen temperatures. As we look at the blue curve, for $k_BT / |J| = 0.1$, the minimum of $\Delta S$ reaches a value of $-0.2462 Nk_B$. This value is smaller in the absolute value compared to the pure system, so when we increase the temperature, the decrease of $\Delta S$ is even more prominent compared to the non-diluted case in Fig. \ref{fig1}.

\section{Conclusions}
We examined how the joint effect of the selective dilution and the external magnetic field can relieve geometrical frustration and how it influences magnetocaloric properties of a selectively diluted Ising antiferromagnet on a triangular lattice. We found out that the interplay between these two factors generally relieves the massive degeneracy in our system and generally reduces negative entropy changes at low temperatures. However, at relatively high temperatures an enhanced magnetocaloric effect was observed at certain selective dilution values. Nevertheless, this interesting finding should be checked by some more reliable technique, such as Monte Carlo simulations.
\section{Acknowledgement}
 This work was supported by the Scientific Grant Agency of Ministry of Education of Slovak Republic (Grant No. 1/0331/15), by the Slovak Research and Development Agency (project APVV-14-0073) and also by the by the Faculty of Science, P.J. \v{S}afarik University (Grant ID. VVGS-PF-2015-490).


\begin{thebibliography}{99}
%
\bibitem{ref1} G.~H. Wannier, \textit{Phys. Rev.} {\bf 79}, 357 (1950). DOI: 10.1103/PhysRev.79.357 \\
\bibitem{ref2} M. Borovsk\'{y}, M. \v{Z}ukovi\v{c}, A. Bob\'{a}k, \textit{Physica A} {\bf 392}, 157 (2013). DOI: 10.1016/j.physa.2012.08.016 \\
\bibitem{ref3} M.~E. Zhitomirsky, \textit{Phys. Rev. B} {\bf 67}, 104421 (2003). DOI: 10.1103/PhysRevB.67.104421 \\
\bibitem{ref4}T. Kaneyoshi, \textit{Acta Phys. Pol. A} {\bf 83}, 703 (1993). DOI: 10.12693/APhysPolA.83.703 \\

%%%%%%%%%%%%%%%%%%%%%
\end{thebibliography}
\end{document}